# Multi-functional 2D hybrid aerogels for gas absorption applications

Charalampos Androulidakis[1], Maria Kotsidi[1,2], George Gorgolis[1], Christos Pavlou[1,2], Labrini Sygellou[1], George Paterakis[1,2], Nick Koutroumanis[1,2] and Costas Galiotis[1,2*]

[1]Institute of Chemical Engineering Sciences, Foundation of Research and Technology-Hellas (FORTH/ICE-HT), Stadiou Street, Platani, Patras, 26504 Greece

[2]Department of Chemical Engineering, University of Patras, Patras 26504 Greece

*Corresponding author: c.galiotis@iceht.forth.gr ; galiotis@chemeng.upatras.gr

## ABSTRACT

Aerogels have attracted significant attention recently due to their ultra-light weight porous structure, mechanical robustness, high electrical conductivity, facile scalability and their use as gas and oil absorbers. Herein, we examine the multi-functional properties of hybrid aerogels consisting of reduced graphene oxide (rGO) integrated with hexagonal boron nitride (hBN) platelets. Using a freeze-drying approach, hybrid aerogels are fabricated by simple mixing with various volume fractions of hBN and rGO up to 0.5/0.5 ratio. The fabrication method is simple, cost effective, scalable and can be extended to other 2D materials combinations. The hybrid rGO/hBN aerogels (HAs) are mechanically robust and highly compressible with mechanical properties similar to those of the pure rGO aerogel. We show that the presence of hBN in the HAs enhances the gas absorption capacities of formaldehyde and water vapour up to ~7 and > 8 times, respectively, as compared to pure rGO aerogel. Moreover, the samples show good recoverability, making them highly efficient materials for gas absorption applications and for the protection of artefacts such as paintings in storage facilities. Finally, even in the presence of large quantity of insulating hBN, the HAs are electrically conductive, extending the potential application spectrum of the proposed hybrids to the field of electro-thermal actuators. The work proposed here paves the way for the design and production of novel 2D materials combinations with tailored multi-functionalities suited for a large variety of modern applications.

**Introduction**

Graphene aerogels (GAs) have attracted significant attention recently since they are ultra-lightweight, mechanically robust, electrically conductive, perform well under harsh environment and, as a result. have found good use in various applications[1]. The GAs can also be used as reinforcing agents in composites, in sensors and nano-electronics, energy storage[2], as cell growth promoters[3], absorbers for the removal of food toxins[4], in catalysis and biomedical applications among others[1]. Other materials such as hexagonal boron nitride (hBN) have also attracted significant attention for the fabrication of aerogels[5] since their distinct properties as compared to GAs, pave the way for the development of new applications. For example, boron nitride based aerogels (BNAs) have been extensively used for the removal of organic and inorganic pollutants from aqueous solutions with high efficiency[6]. Moreover, BNAs possess high thermal conductivity and they remain stable up to $1000°C$ which make them attractive materials for applications required to perform reliably under harsh environments[7]. Besides graphene and boron nitride based aerogels, the area is highly active and other materials have also been tested, including silica[8,9] and $SnO_2$[10] based aerogels that show attractive properties and have been used in various applications.

In an attempt to combine and exploit the distinct properties of dissimilar 2D materials, the fabrication of 2D hybrids could, under certain circumstances, impart new multi-functionalities and therefore has recently emerged as a very intense research field[11-17]. In this regard, hybrid aerogels (HAs) have started to gain attention and some efforts have already been reported in the literature for aerogels composed of combined graphene/boron nitride nano-sheets[11,12], graphene/ boron nitride nanotubes[13], graphene/$MoS_2$[14-16] and others[17]. In particular, BNAs exhibit high thermal conductivity but low electrical conductivity[18] therefore, they are only suitable for applications as insulators. In contrast, reduced graphene oxide has both high thermal and moderate electrical conductivity and therefore its field of application differs than hBN. Both materials possess attractive mechanical properties, they are highly compressible and perform well as absorbers of gas and oil pollutants. It would be of great interest to create hybrid aerogels that retain the properties of the standalone materials or even develop hybrids that could outperform the efficiency of the homostructured materials.

One particular application of the aerogels is their use as volatile organic compounds (VOC) absorbers for the protection of museum artefacts such as paintings and other objects. It is worth noting that most artefacts owned by museums are often stored under hostile environmental conditions in deposits (e.g. in crates and boxes). It has been estimated that the largest museums typically display only about 5% of their collection at any time. This roughly 5% is usually rotated among the most important works the museums have, while other less significant works may never leave the storage area. Additionally, some artworks may be displayed but need preservation frequently in order that they survive, like works based on paper, which will fade if exposed to light. Both carbon and hBN aerogels have been tested as oil and organic absorbers[19]. Bi et al.[20] examined the efficiency of spongy graphene for oil absorption by floating the sample in liquid pollutants such as commercial petroleum products, fats and ketones, and the results showed high speed of absorption and efficient recovery. Similar experiments performed for thermally cross-linked Poly(acrylic acid)/rGO of low porosity as absorbers of various oils with high efficiency[21], as well for pure graphene aerogels with density less than ~9 mg/cm$^3$ [22]. An extensive discussion on the subject regarding carbon based absorbers can be found in reference[19] and for hBN based materials in reference[23]. Boron nitride based porous materials have also been tested for the removal of substances from water. A foam like hBN structure was developed with density of ~30.4 mg/cm$^3$ and proved very efficient for the removal/separation of a number of oils and chemicals by immersing the porous materials into contaminated water[24]. Similar experiments performed for another sponge like material made by few-layer hBN-coated melamine sponge which showed to possess very good recyclability performance[25].

Herein, we fabricate hybrid aerogels at various mix ratios composed of reduced graphene oxide (rGO) and hexagonal boron nitride (hBN) platelets using a freeze-drying approach. The method is cost effective and can be easily used for the fabrication of large-scale quantities, while it can be extended to the fabrication of other 2D materials as well. In this work we examine in detail the mechanical properties of the HAs as well the changes in the electrical conductivity induced by the presence of hBN. Furthermore, we investigate the efficiency of the proposed HAs as gas absorbers of formaldehyde and hydrochloric acid, as well their capacity for water vapour absorption. The results presented below show that the rGO/ hBN hybrid aerogels possess good mechanical properties and reasonable electrical conductivities but, most importantly, certain combinations of these hybrids exhibit extremely high gas absorption capabilities.

**Experimental**

The approach presented previously by Hong et al.[26] was followed herein for the fabrication of the rGO aerogels. Aqueous solution of GO was prepared by modified Hummer's method[27,28] and was subsequently diluted in water to obtain a concentration of 1 mg/mL. Hypophosphorous acid ($H_3PO_2$) and iodine ($I_2$) of weight ratio GO:$H_3PO_2$:$I_2$ is 1:100:10, were then added as the chemically reducing agents[29]. Subsequently the solution placed in a furnace and was heated to 80°C for 8 hours, resulting in a uniform gelation of the GO. The sample was then rinsed with water until a pH of equal to 5, followed by freeze-drying for 48 h. For the fabrication of the HAs, another solution with boron nitride platelets of 1 mg/mL concentration was prepared in ethanol-water (50-50 % w/w) and stirred for 10 minutes. Solutions with mixed GO and hBN with various concentrations of 0.9/0.1, 0.7/0.3 and 0.5/0.5 mg/mL were stirred and then the same steps as for the case of the pure GO were followed. We note that the addition of ethanol significantly stabilizes the mixed solution[30] and assists to achieve uniform dispersions as seen in figure 1b. When we used only water, we observed that the hBN material tends to separate from the mixed solution and even though gelation still occurs, there is no uniform distribution of the hBN platelets in the formed aerogel. This bottleneck is attributed to the absence of polar groups to the surface of hBN[13]. It is noted that this approach cannot be used for the preparation of pure hBN aerogels, which points out to the crucial role of the presence of the GO which stabilizes the mixture. This is due the gelation mechanism that involves the simultaneous reduction and self-assembly of the GO sheets, based on chemical reactions between the oxygen functional groups present at the surface of the GO and the chemical agents[31]. Thus, the hBN platelets that do not contain such groups on their surface and simply attach to the GO sheets. The only complication is that the gelation time might be increased in the mixed solutions as discussed previously[32]. The density of the aerogels is found to be in the range of 13.9-23.4 mg/cm$^{-3}$ without significant variation between samples.

The surface analysis measurements were performed in a UHV chamber (P~5×10−10 mbar) equipped with a SPECS Phoibos 100-1D-DLD hemispherical electron analyser and a non-monochromatized dual-anode Mg/Al x-ray source for XPS. The XP Spectra were recorded with MgKa at 1253.6 eV photon energy and an analyser pass energy of 10 eV giving a Full Width at Half Maximum (FWHM) of 0.85 eV for Ag3d$_{5/2}$ line. The analysed area was a spot with 3 mm diameter. The atomic ratios were calculated from the intensity (peak area) of the XPS peaks weighted with the corresponding relative sensitivity factors (RSF) derived from the Scofield cross-section taking into account the electron transport properties of the matrix, (namely the inelastic mean free path (IMFP) λi and the elastic-scattering correction factor Q, depend mainly on the corresponding electron kinetic energy (KE)) and the energy analyser transmission function. For spectra collection and treatment, including fitting, the commercial software SpecsLab Prodigy (by Specs GmbH, Berlin) was used. The XPS peaks were deconvoluted with a sum Gaussian-Lorenzian peaks after a Shirtey type background subtraction.

The gas absorption test was conducted in static conditions into a closed glass desiccator with an excess of pollutant, using a saturated vapour stream at room temperature. 100 ml of either formaldehyde (37 wt.% in $H_2O$) or hydrochloric acid were used as gas source each time. For the humidity measurements, veils full of water were placed similarly to the gases, while the relative humidity (RH) was measured independently and was found to be ~97%. All samples were initially dried at 200°C for 2h in order to remove their absorbed humidity and weighed into a high accuracy weight meter to measure their dried mass. Afterwards, the aerogels were loaded in a glass petri dish which was mounted inside the desiccator. The desiccator with the aerogels and the fuming gas was constantly inside a fume hood. Periodically, the mass of each sample was recorded by the weight meter, exactly next to the fume hood minimizing the exposure of the aerogels to the environment.

The regeneration of the graphene and the graphene-hBN aerogels was performed by using a common electric hair dryer. During the regeneration process, the formed volatile vapour was removed with the assistance of the hood, as reported by Li et al.[33].

**Results and Discussion**

The approach presented previously by Hong et al.[26] was followed herein for the fabrication of graphene based rGO aerogels and is described in detail in the experimental section. For the fabrication of the HAs, another solution with boron nitride platelets was prepared in ethanol-water and mixed with the GO in various concentrations of 0.9/0.1, 0.7/0.3 and 0.5/0.5 mg/mL, and the same procedure as in the case of the GA was pursued. We note that the addition of ethanol significantly stabilizes the mixed solution[30] and assists to achieve uniform dispersions as seen in figure 1b. The density of the aerogels is found to be in the range of 13.9-23.4 mg/cm$^{-3}$ without significant variation between samples. The advantages of the present approach for the fabrication of hybrid aerogels are that it can be easily used for scalable fabrication since it is less energy demanding than other approaches[34]. It must be stressed that this synthesis method cannot be used for the fabrication of pure hBN aerogels.

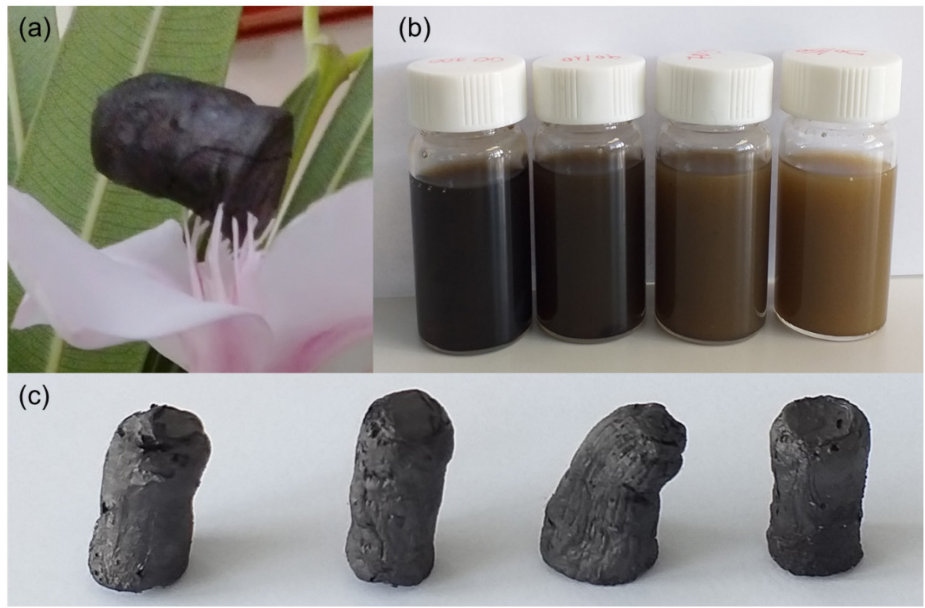

**Figure 1**. a) An ultra-light weight rGO aerogel which is shown to be supported by just the stamen of a flower inside the lab, b) starting solutions for the development of the HAs: from left to right, neat rGO, rGO-hBN 90/10, rGO-hBN 70/30 and rGO-hBN 50/50 and c) the as-prepared hybrid aerogels with different amounts of graphene oxide and hBN with the same order as in (b).

In figure 1, the as-prepared hybrid aerogels are shown and the corresponding solutions prior to gelation and subsequent freeze-drying. A variety of techniques was employed for the characterization of the HAs including X-ray photoelectron spectroscopy (XPS), X-ray diffraction (XRD), Raman spectroscopy and scanning electron microscopy (SEM). The Survey XPS Scans (figure 2a) show the presence of C, O and P atoms in both pure GA and HA samples, while the trace of N and B is present on the HAs sample surface. The deconvoluted XPS C1s is shown in figure 2b,c. By analysing the peak (Table 1) we detect the presence of $sp^2$ and $sp^3$ hydridization and also of oxides in the form of epoxides, hydroxides, carbonyls and carboxyls due to pi-pi* transition loss peak[35]. From the peak areas of the C1s, the percentage (%) component concentration is calculated and from the peak intensities of C1s, O1s, P2p and N1s the (%) relative atomic ratio is obtained. Both sets of results are shown in Table 1. In order to calculate the C:O atomic ratio in each specimen, the oxygen concentration due to $P_2O_5$ chemical state is subtracted. The results are for the rGO-hBN (50/50) C:O=5.6, and C:O=10 for the pure GA, suggesting the effective reduction of the GO to rGO[36]. The C:O ratio for the 50/50 is almost the half of the pure GA which shows that the presence of hBN affects the reduction of the GO.

**Table 1**: Percentage of C1s component concentration derived from the C1s peak deconvolution (figure 2) and relative atomic ratio C:O:P of the GA and C:O:P:N:B of the HA sample.

| sample | % component of C bonds | | | | | | Relative atomic ratio |
|---|---|---|---|---|---|---|---|
| | C-C sp2 | C-C sp3 | C-O(H) | C=O | COOH | pi-pi* | C:O:P:N:B |
| HPO | 68,9 | 5,7 | 12,8 | 5,4 | 4,2 | 3,0 | 1:0.4:0.12 |
| hBN | 71,1 | 9,1 | 12,8 | 2,9 | 2,1 | 2,0 | 1:0.34:0.065:0.031:0.029 |

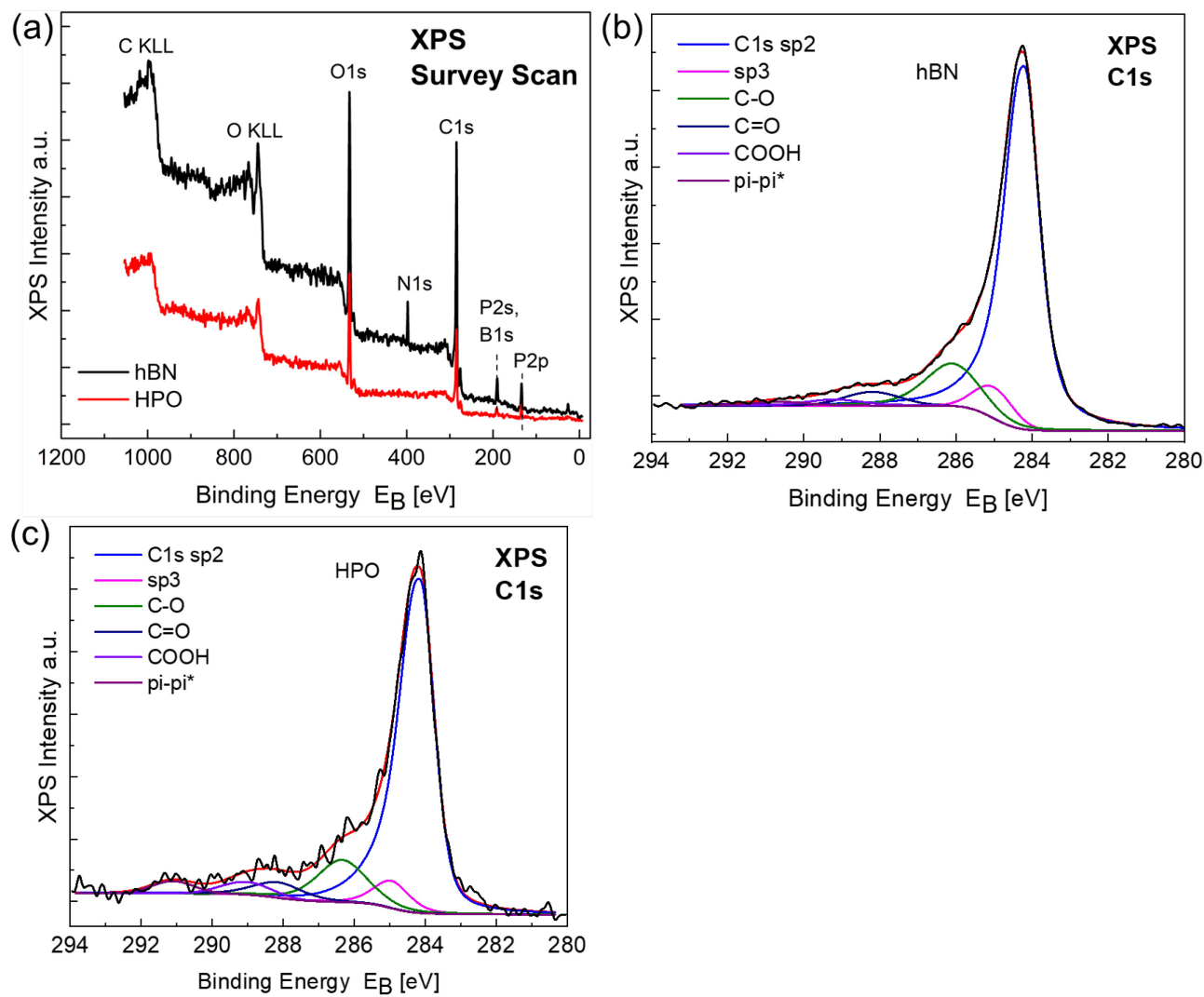

**Figure 2**. (a) Survey Scans of the hBN and HPO samples. (b, c) Deconvoluted C1s XPS peaks of the hBN and HPO samples, respectively.

In figure 3 Raman and XRD measurements are shown for the rGO and HA samples. Both the Raman and XRD measurements prove the presence of both materials in the hybrid aerogels. The XRD plots show a broad peak at $2\theta = 26°$ which corresponds to the (002) plane of graphite structure, as reported elsewhere[37]. The HAs aerogel show graphite peak at $2\theta = 26°$ and also a clear peak at 27.3°, which are attributed to the (002) diffraction of hBN[12]. The position of the Raman peaks for the GO are ~1590 cm$^{-1}$ and 1356 cm$^{-1}$ corresponding to the G and D phonons, respectively[38]. The presence of hBN is also evident in the HA from an intense peak at 1366 cm$^{-1}$ originating from the in-plane $E_{2g}$ phonon mode[39]. The intensity of the peak of the hBN is large due to the relatively large thickness (~2-3 μm) of the platelets.

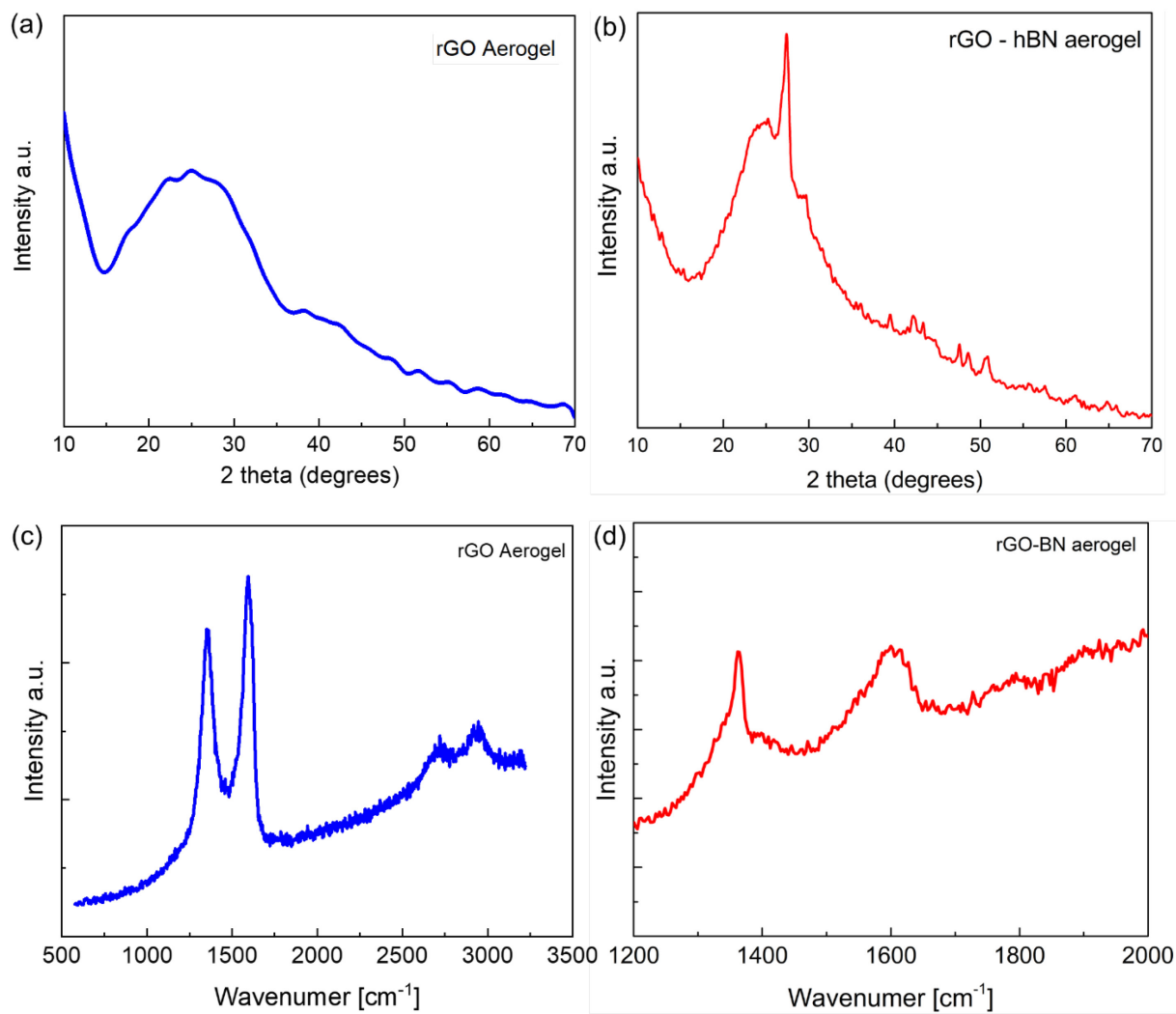

**Figure 3**. (a), (c) XRD and Raman spectra of the rGO and the corresponding spectra (b), (d) for the rGO-hBN aerogels samples. The as-made rGO aerogel exhibits a weak broad peak, characteristic of an amorphous material, at 2θ = 26°, which corresponds to the (002) plane of graphite structure, as reported elsewhere[37]. The hybrid rGO – hBN aerogel shows mainly the before-mentioned rGO peak at 2θ = 26° and also a clear peak at 27.3° which is attributed to the (002) diffraction peak of hBN[12]. For characteristics of the G and D peaks are clearly seen in the Raman spectra for the rGO and also the presence of hBN is confirmed by the peak of the $E_{2g}$ phonon at ~1366 cm$^{-1}$.

SEM images for all samples with various mixed ratios are given in figure 4. The hBN platelets are clearly distinguished and seem to be either attached to the surface of the rGO layers or wrapped around them, creating a robust macro-scale porous network. We also observe from the SEM images that the magnitude of the pores tends to increase with the increase in the amount of hBN in the HA samples. This can be explained by considering the amount of the rGO in each structure and the relative volume of the resulted aerogel. The 3D porous assembled structure depends on the available amount of rGO. The resulted aerogels show similar macroscale volumes in all cases. Thus, with decreasing rGO, less amount of material assembles and occupies the same volume (since the hBN is wrapped around the rGO and does not contribute to the formation of the 3D network), as seen in the optical pictures presented in figure 1. Moreover, this is also manifested in the average values of the density given in Table 2 which is less than 30 mg/cm$^{-3}$ for all cases. In fact, this value is a prerequisite for classifying the 3D porous structure as an aerogel[12]. The structural characteristics of the integration of the rGO with the hBN, such as the attachment and wrapping of the hBN to the rGO, agrees very well with the results reported in other studies[11,12].

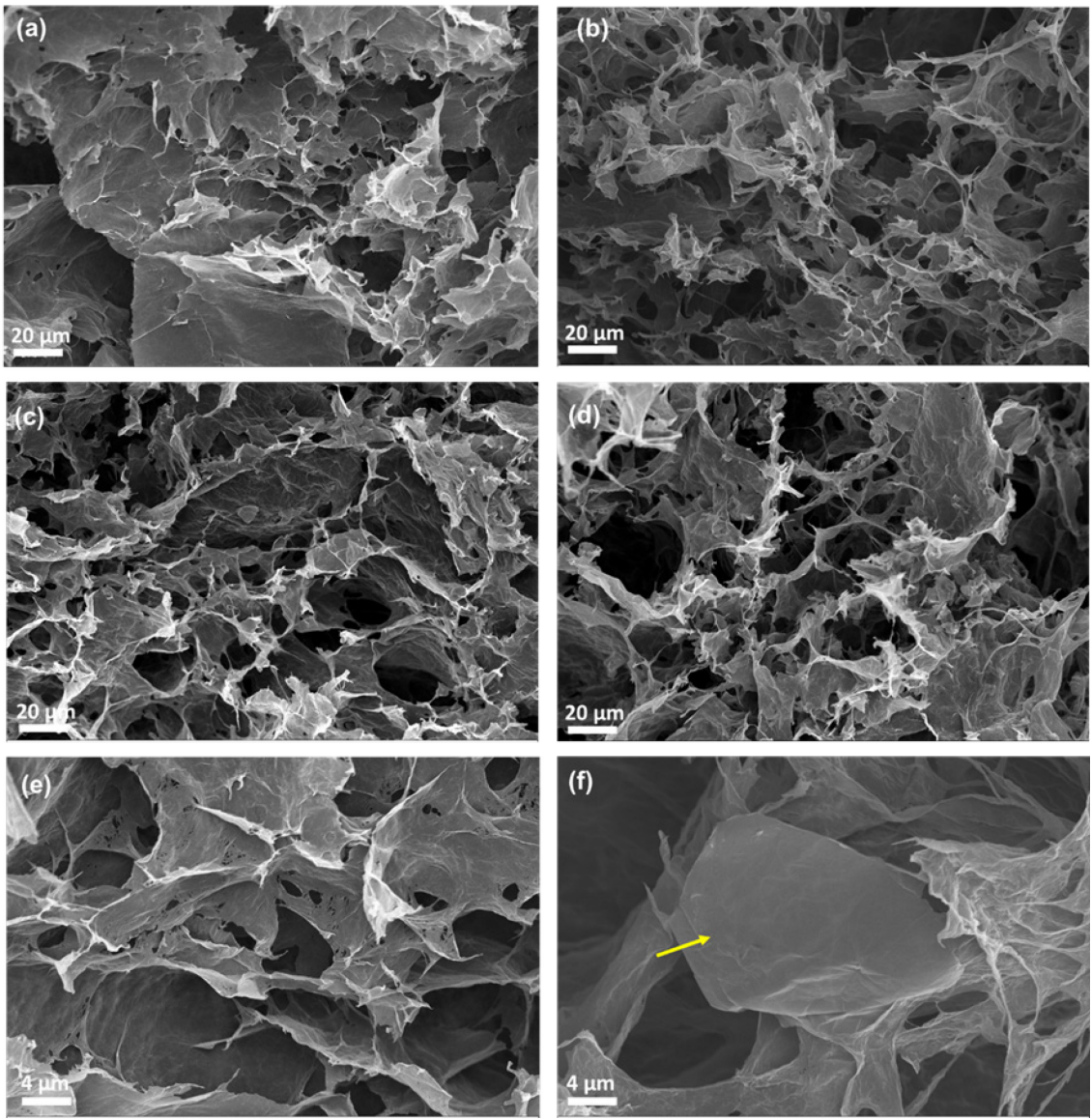

**Figure 4.** SEM images of the (a) rGO aerogel, (b) 90/10 HA, (c) 70/30 HA and (d) 50/50 HA. In (e) a zoom of the rGO aerogel is presented and in (f) a zoon of hybrid aerogel which clearly show the presence of hBN platelets in the HA samples and have size of a few microns in length and about ~3-5 microns thickness. The scale bar is 20 μm for (a-d) while for (e, f) is 4 μm.

The mechanical properties under compression for the hybrid structures of variable rGO/ hBN content were examined in detail and compared to those of the pure rGO aerogel. In figure 5a representative stress-strain curves for all tested samples are presented and in figure 5b the modulus along with the electrical conductivity of all cases. All samples compressed up to a maximum compressive strain of ~80-90% without failure, followed by recording the unloading behaviour. The stress-strain response consists of three discrete stages, characteristic of the compressive behaviour of aerogels. Initially a linear Hookean behaviour, which holds roughly up to 10%, is observed corresponding to the elastic bending of the cell walls. For all cases the linear stiffness is similar with values in the range of ~64-80 kPa. The initial linear behaviour from which the Young's modulus is extracted, is followed by a non-linear regime of much lower modulus. During this regime, densification of the compressed walls of the aerogel sets off, until a compressive strain of 60−70%. For higher compression, the stress increases rapidly due to the resulting densification and closing of the pores of the 3D structure. Some differences in the compressive behaviour are plausibly expected due to the non-perfect cylindrical geometry of the samples and the variation in the density, but overall, the results are consistent within the experimental error. The differences in the energy dissipation are also attributed to the variation in the density of the samples. We note that the similar mechanical properties (maximum compressive stresses and modulus) of the present HAs with the pure GA reflects the efficiency of the fabrication procedure and the effective synergy of the two materials. It seems therefore that using rGO to support mechanically the hBN is a very effective strategy also for fabrication purposes, since high aspect ratio HAs can be easily produced and more importantly the distinct properties of both materials can be preserved as discussed below.

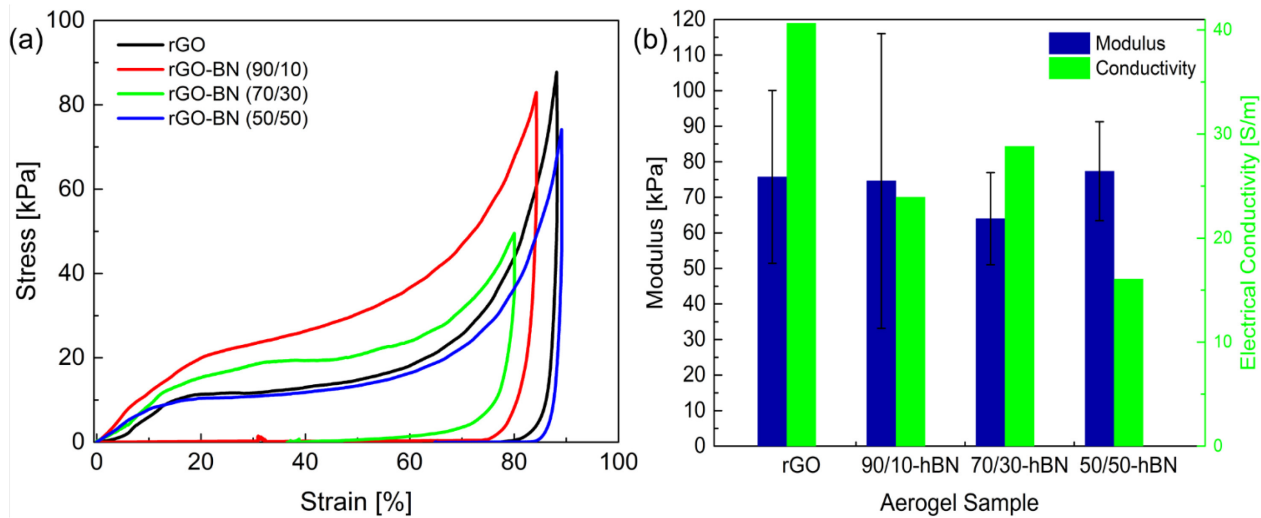

**Figure 5**. (a) Representative stress-strain curves under compression for all the examined aerogels, namely rGO, 90/10, 70/30 and 50/50. (b) Compressive modulus (left) and electrical conductivity (right) for all tested aerogel samples.

Table 2. The average density of the samples for all cases. In the second column the density of specific samples for which the electrical conductivity was measured and is given the last column. The error values represent the standard deviation of the statistical analysis applied on the measured densities of the samples.

| Sample | Average Density [mg/cm$^3$] | Density [mg/cm$^3$] | Electrical Conductivity [S/m] |
|---|---|---|---|
| GA | 18.1 ± 2.1 | 19.4 | 40.4 |
| HA-90/10 | 15.2 ± 2.2 | 11.9 | 23.3 |
| HA-70/30 | 20.2 ± 4.2 | 23.4 | 28.1 |
| HA-50/50 | 17.5 ± 1.6 | 18.1 | 15.6 |

In a previous study, similar HAs with a smaller density of ~3.6 mg/cm$^{-3}$ were tested and a compressive stress of ~1.5 kPa was measured at 50% of strain[11]. Elsewhere, rGO aerogels reinforced with boron nitride nanotubes of total density ~16 mg/cm$^{-3}$, reached stress of ~12 kPa at ~75% of compressive strain[13]. We note that the mechanical properties of aerogels are linearly related to the density of the samples[40,41], and the modulus for the GA of the present study agrees well with the analysis of reference[40].

Thus, the present HAs possess better mechanical performance as compared to reported results for other heterostructures[11,13] since they have similar density or slightly higher, and in our case they reach much higher compressive stresses without failure.

In figure 4b, the electrical conductivity for all cases is presented. In Table 1 the electrical conductivity and the density of the corresponding specimens measured are given. The neat rGO aerogel presents a relatively high conductivity with value of ~40.4 S/m, indicating the effective reduction of the GO in agreement with the results from the XPS measurements. This value is close to the highest reported conductivities in the range of 50-87 S/m for graphene aerogels with similar densities[40,42]. The 90/10 hybrid exhibits almost half of this value of ~23.3 S/m, and with further increment of the amount of the hBN the conductivities do not vary a lot within the experimental error (Table 1). Overall, any differences are attributed to density variation, the extend of the reduction of the rGO, as well as, of course the significant differences in electrical conductivity between the two constituents. Effectively, this means that the conductivity of the pure GA and the 50/50 sample should differ by a factor of 2, assuming they possess the same density. Instead, from table 1 we observe that despite these two samples have similar density, the conductivity of 50/50 is smaller by a factor of 2.6. This additional decrease is due to the less effective reduction of the GO to rGO with the presence of hBN as evident by the XPS measurements. At any rate, the electrical conductivity is considered reasonably high for all cases. The electrical conductivity along with the excellent thermal behaviour and stability of hBN at high temperatures[12,43], can be exploited in a variety of applications for these aerogels, for example as electro-thermal harvesters[44,45]. The results are summarized in figure 5b.

Both GAs and BNAs are highly efficient gas and oil absorbers[25,46,47], thus we examine the capacity of the present aerogels for their use in gas absorption applications. We note that a potential application is the use of the proposed aerogels for the protection from volatile organic compounds (VOCs) and humidity of important artefacts in storage facilities (the number surpass the 10000) which is one of the main objectives of the present work.

We tested the efficiency of the proposed hybrids and the GA for absorption of formaldehyde and hydrochloric acid, which are among the most commonly examined gases, as well the water vapour (humidity) absorption. Formaldehyde, in particular, is considered as a model polar VOC pollutant[48]. Prior to exposing the samples to the gas/humid environment, care was taken to dry the samples by heating to 200° C, followed by immediate measurements of their initial weight under dry conditions. The samples were then placed in a desiccator and exposed to saturated gas controlled environment, while their mass weight gain due to absorption was monitored by regular gravimetric measurements until the maximum mass increment was reached. More details can be found in the experimental section. From figure 6 it is seen that as the concentration of hBN into the hybrid aerogels increases, so does the maximum absorption of formaldehyde. Thus, the most efficient absorber of formaldehyde is the hybrid aerogel with the highest amount of hBN, the 50-50 sample. We can observe that the 90-10, 70-30 and 50-50 samples show maximum absorptions higher by 3, 6 and 7 times in comparison to the neat rGO sample. This proves the capability of hBN in absorbing formaldehyde, and to the best of our knowledge, this is the first time that such a structure, like a hybrid aerogel from rGO and hBN, is reported to exhibit this behaviour. On the contrary, for the case of hydrochloric acid the opposite trend is observed with the rGO sample being the most efficient absorber. The absorption is slightly reduced for the 90-10 sample, while higher decrease occurs for the other cases.

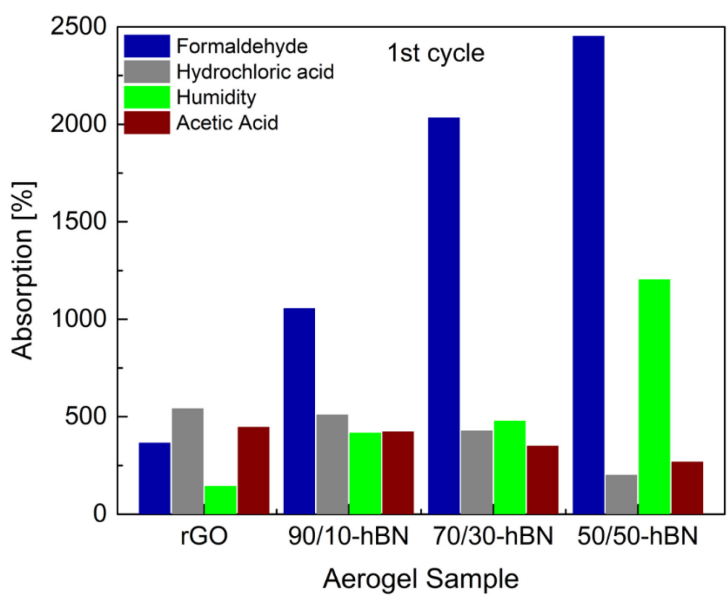

**Figure 6**. Absorption capacity of the rGO and hybrid rGO-hBN aerogels for formaldehyde, hydrochloric acid, humidity and acetic acid.

The high absorption of formaldehyde of the HAs can be mainly ascribed to the polarity of the molecules of formaldehyde. It is considered that materials with high specific surface area and a hydrophilic surface can be effective adsorbents for formaldehyde[49]. Formaldehyde is physiosorbed on graphene[50,51], and this absorption is weak because of its small binding energy, large binding distance and small charge transfer from formaldehyde to the graphene sheets[51]. On the other hand, hBN (sponge like) exhibits superb formaldehyde absorption performance due to its high specific surface area and its abundant surface hydroxyls and amines[49]. Furthermore, the HAs are more hydrophilic compared to the rGO as deduced form the humidity absorption measurements discussed below, which also contribute to the high performance toward gaseous formaldehyde[49]. Additionally, chemisorption takes place on the hBN layers and contributes to high formaldehyde absorption via Cannizzaro-type disproportionation reactions, during which formaldehyde is converted into less toxic formic acid and methanol[49]. Another factor that can enhance the formaldehyde absorption is the surface properties of hBN. Hydrogen bonding between formaldehyde and hydroxyl or amine groups is very crucial in the absorption of formaldehyde[52,53], while the formaldehyde molecules and hBN layers show both planar configuration, resulting in lower adsorption resistance[49]. Finally, formaldehyde has π-bonding between C and O atoms, while hBN also shows large 2D delocalized π-bonded structure. Consequently, π–π bonding interaction can occur between hBN and formaldehyde and this strong interaction enhances the absorption of the latter[49].

As already mentioned, formaldehyde is known to be weakly physiosorbed on reduced graphene oxide sheets as a consequence of its small binding energy to graphene which is equal to $E_b = 0.29$ eV[54]. When the formaldehyde molecules are adsorbed onto the neat rGO sheets, formaldehyde shows the preference to be located perpendicular to the sheet with H atoms close to the sheet, as shown in Figure 7a below. This results to an angle of 90° between the sheet and the axis passing through C=O bond of formaldehyde. Also, when formaldehyde is adsorbed onto the rGO sheets, creates an energy band gap at the Dirac point of graphene[54,55]. Thus, graphene then becomes a n-type semiconductor, the valence electrons are increased and charge transfer from the electron donor molecules, formaldehyde, to the graphene sheets takes place[54].

As the amount of hexagonal boron nitride is gradually increased in the hetero-structured aerogel, the adsorption of formaldehyde increases (from 90/10 % wt. to 50/50 % wt.). This can be explained due to the strong chemisorption between the molecules of (adsorbed) formaldehyde and the 'doped' with hBN, graphene. It has been shown that doping graphene with boron (B) and nitrogen (N) atoms enhances the interaction between the graphene and formaldehyde significantly[56]. Formaldehyde exhibits larger adsorption energy and net charge transfer when is chemisorbed on doped graphene in comparison to a corresponding undoped graphene sheet. Specifically, for a boron-doped graphene sheet, the adsorption energy ($E_{ads}$) and charge transfer (Q) are equal to −0.543 eV and 0.239 e respectively, while the corresponding values for pristine graphene are −0.137 eV and 0.094 eV, respectively[56]. This high adsorption energy is a strong indication that chemical bonds are created between the dopants (B, N) and O atom of formaldehyde. The reason behind this behaviour is the hybridisation between the dopants (B, N) and C atoms, around the two dopant atoms. Especially for boron, an overlap between B-2p and C-2p orbitals, with the energy ranging from −3.810 to −2.236 eV is observed[56], which proves the appearance of a strong hybridization between B and C atoms around the doping atom. This strong orbital interaction between the p orbitals of B, N, and the C atoms also alters the magnetic properties of such systems[56].

For the interaction between the boron nitride platelets and the formaldehyde, Cannizzaro-type disproportionation reaction of the latter has been reported to be dominant[57]. The related mechanism is shown in Figure 7b. During this reaction, the molecules of formaldehyde are adsorbed onto the boron nitride. The nitrogen atoms of the boron nitride sheet function as a Lewis base and initiate a reaction of nucleophilic addition during which a hydrogen atom of formaldehyde is transferred to another similar molecule (phase 1). Consequently, adsorbed formaldehyde molecules are transformed into HCONH2 (formamide) and others into methoxy salts (phase 2). Afterwards, the produced formamide and methoxy salts react with water adsorbed onto the surface of the material to produce in turn less toxic formic acid and methanol (phase 3). The last phase indicates the critical role of the hydrophilicity for the examined hybrid aerogel.

Regarding the adsorption of hydrochloric acid onto the examined pure rGO and hybrid aerogels, a weaker interaction between the HCl molecules and the surface of the prepared materials can be observed, as shown in Figure 6. Hydrochloric acid is physiosorbed by graphene[58,59] as can be deduced by the same approximately adsorption rates of all the hybrid aerogels that were examined in HCl, and neat rGO inside formaldehyde. It has been found that especially for a single chlorine atom, the bonding is ionic through the transfer of charge from graphene to chlorine adatom and creates a small local distortion in the underneath planar graphene[58]. The transfer of a single chlorine adatom on the surface of graphene is achieved almost without barrier. But, the anchoring of a graphene surface with Cl adatoms resulting in several conformations cannot be maintained because of the strong Cl-Cl interaction which in turn facilitates the desorption through the creation of Cl2 molecules.

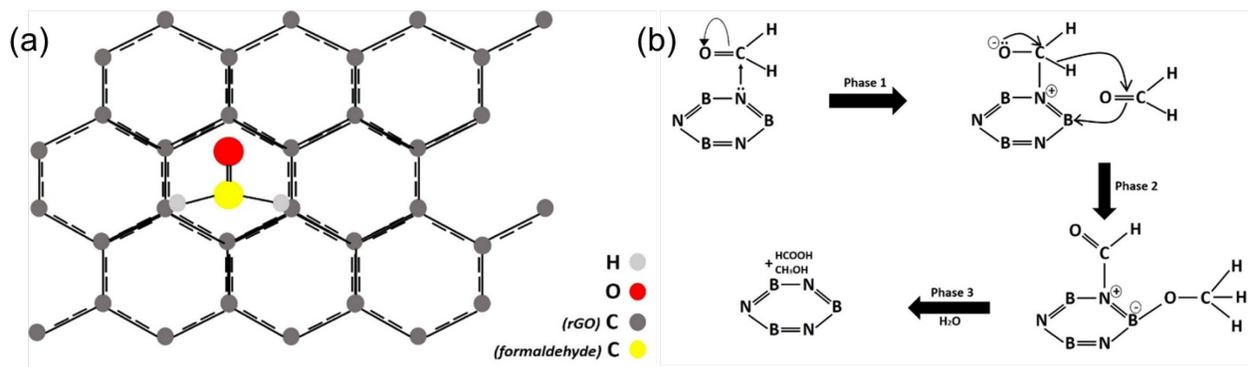

**Figure 7**. (a) Adsorption configuration of formaldehyde on reduced graphene oxide (rGO) sheet. (b) Possible disproportionation reaction for formaldehyde over the surface of porous boron nitride.

The absorption capacity for acetic acid is a bit smaller as compared to the hydrochloric acid but a similar trend holds with decreasing absorption with increased amount of hBN. Similar experiments have been performed for testing reduced graphene oxide based porous materials as gas absorbers for various VOCs including benzene, toluene, carbon dioxide and formaldehyde[60-66]. For comparison they are listed in the supporting information (Table S1) including the results of the present work, along with their gravimetric measurements and details on the structure of the examined samples.

The capability of moisture absorption of the HAs was also examined after their insertion into a closed plastic chamber in which vials full of water were placed inside as the source of humidity. The relative humidity (RH) was measured to be ~97% by the addition of a humidity sensor inside the chamber. Following the same procedure with the gas absorption measurements, the samples were firstly dried at 200° C and then being weighed periodically until they achieved a maximum absorption. As shown in figure 6, the addition of hBN platelets increases significantly the humidity absorption which can be up to ~8 times higher than the absorption of the neat rGO sample. The increment reaches a value of ~1203.7% for the 50/50 HA sample of its initial weight. These results show that the hBN induces a highly hydrophilic nature to the HAs, in agreement with recent experiments on the wettability properties of pure hBN aerogels[67]. In Tables 3-5 we present the initial weight of the samples before subjecting them to the gas or humid environment, as well their weight after drying for removing the absorbed gas/humidity. The samples fully recover to their initial weight, exhibiting an excellent regeneration which a prerequisite for their effective use as gas absorbers. We also tested the influence of humidity in the absorption capacity for each examined VOC. We measured the humidity level inside each desiccator, and we performed independent measurements for humidity absorption at these RH levels. The RH inside the desiccators found to be in the range 63-75.5% with the upper value corresponding to the formaldehyde. From the independent RH measurements, we observed that the samples begin to gain weight due to humidity when the RH overcomes the value of RH~70%. Close to the upper limit of RH~75%, the corresponding percentage increase of the aerogel weight was found to be only 0.22% and it was lower for RH<70%. Thus, we can safely deduce that the influence of humidity in the VOC absorption is negligible.

As already mentioned, we cannot fabricate pure hBN aerogels with this approach, thus we cannot have a direct comparison for this case. The absorption results show that the two materials have different efficiencies depending on the specific vapour. The absorption efficiency is not compromised by mixing of the 2D materials, thus, in order to make as efficient absorbers as possible for a wide range of vapour concentrations, we need to use more than one material. Given the large availability of 2D materials with diverse properties, there is much room for the fabrication of novel highly efficient absorbers beyond the state-of-the-art. Moreover, these hybrids can be exploited in other applications for which electrical and thermal conductivities are required.

**Conclusions**

In summary, we presented a facile fabrication route for 2D hybrid aerogels comprising of rGO mixed with hBN platelets. The fabrication method is cost effective and scalable compared to high energy demanding methods for the fabrication of pure hBN aerogels, and can be used for mixing other 2D materials with rGO. The resulted HAs are mechanically robust with mechanical properties similar to those of pure GA. The HAs are electrically conductive, despite a decrease that occurs compared to the pure rGO aerogel, which can be useful for applications for which electrical conductivity is required. The most striking result is the very effective absorption capacity of the HAs of formaldehyde and humidity, which is 7-8 times higher compared to the pure GA and at least one order of magnitude higher than a pure hBN aerogel[49]. Furthermore, the addition of hBN provides superior thermal conductivity to the HAs[12], which make them promising materials for applications at elevated temperatures, further expanding their potential use. The HAs can be used as more efficient VOC absorbers for absorption of multiple compounds, in contrast to the standalone materials which are not efficient for all cases. All the above give the HAs a high degree of multi-functionality which enables their use in a plethora of applications, and, as demonstrated in the main text, the proposed aerogels can be used for protection of artefacts such as paintings in storage facilities in which a high concentration of humidity and VOCs is encountered.

**Table 3: SAMPLES WITH FORMALDEHYDE**

| Aerogel sample | Mass after drying at 200° for 2 h | Mass after drying with the hair dryer for |
|---|---|---|

| | [mg] | 24 h [mg] |
|---|---|---|
| 100% Graphene | 11.3 | 11 |
| 90% Graphene-10% hBN | 15.7 | 15.6 |
| 70% Graphene-30% hBN | 18.6 | 17.5 |
| 50% Graphene-50% hBN | 16.1 | 15 |

**Table 4: SAMPLES WITH HYDROCHLORIC ACID**

| Aerogel sample | Mass after drying at 200° for 2 h [mg] | Mass after drying with the hair dryer for 24 h [mg] |
|---|---|---|
| 100% Graphene | 16.7 | 17.7 |
| 90% Graphene-10% hBN | 19.7 | 17.0 |
| 70% Graphene-30% hBN | 18.3 | 18.2 |
| 50% Graphene-50% hBN | 18.2 | 18.5 |

**Table 5: HUMIDITY ABSORPTION**

| Aerogel sample | Mass after drying at 200° for 2 h [mg] | Mass after drying with the hair dryer for 24 h [mg] |
|---|---|---|
| 100% Graphene | 17.7 | 17.8 |
| 90% Graphene-10% hBN | 17 | 16.7 |
| 70% Graphene-30% hBN | 18.2 | 18.6 |
| 50% Graphene-50% hBN | 18.5 | 16.2 |

**Acknowledgments**

The authors acknowledge and thank the Laboratory of Electron Microscopy and Microanalysis (L.E.M.M.) located at the Department of Biology at University of Patras for providing us access to their freeze-drying instrument. Dr Vassileios Dracopoulos is thanked for assisting on the SEM measurements. Ms Maria

Smyrnioti is also thanked for her assistance in performing nitrogen (N2) sorption experiments. CA, MK, GG and CG acknowledge the support from "APACHE", Active & intelligent Packaging materials and display cases as a tool for preventive conservation of Cultural Heritage" which is implemented under the EU-Horizon 2020. CG, GP and CP acknowledge the support from "Graphene Flagship Core Project 3", SGA: 881603 which is implemented under the EU-Horizon 2020 Research & Innovation Actions (RIA) and is financially supported by EC-financed parts of the Graphene Flagship.**References**

1  Gorgolis, G. & Galiotis, C. Graphene aerogels: a review. *2D Materials* **4**, 032001 (2017).
2  Myung, Y. *et al.* Graphene-based aerogels derived from biomass for energy storage and environmental remediation.  **7**, 3772-3782 (2019).
3  Shukla, S. *et al.* Sustainable graphene aerogel as an ecofriendly cell growth promoter and highly efficient adsorbent for histamine from red wine.  **11**, 18165-18177 (2019).
4  Bajpai, V. K. *et al.* A sustainable graphene aerogel capable of the adsorptive elimination of biogenic amines and bacteria from soy sauce and highly efficient cell proliferation.  **11**, 43949-43963 (2019).
5  Lei, W. *et al.* Boron nitride colloidal solutions, ultralight aerogels and freestanding membranes through one-step exfoliation and functionalization. *Nature communications* **6**, 8849 (2015).
6  Yu, S. *et al.* Boron nitride-based materials for the removal of pollutants from aqueous solutions: a review. *Chemical Engineering Journal* **333**, 343-360 (2018).
7  Li, G. *et al.* Boron Nitride Aerogels with Super-Flexibility Ranging from Liquid Nitrogen Temperature to 1000° C. *Advanced Functional Materials* **29**, 1900188 (2019).
8  Parale, V. G., Lee, K.-Y. & Park, H.-H. J. J. o. t. K. C. S. Flexible and transparent silica aerogels: an overview.  **54**, 184-199 (2017).
9  Patil, S. P., Parale, V. G., Park, H.-H., Markert, B. J. M. S. & A, E. Molecular dynamics and experimental studies of nanoindentation on nanoporous silica aerogels.  **742**, 344-352 (2019).
10  Parale, V. G. *et al.* SnO2 aerogel deposited onto polymer-derived carbon foam for environmental remediation.  **287**, 110990 (2019).
11  Li, H. *et al.* Multifunctional and highly compressive cross-linker-free sponge based on reduced graphene oxide and boron nitride nanosheets. *Chemical Engineering Journal* **328**, 825-833 (2017).
12  An, F. *et al.* Highly anisotropic graphene/boron nitride hybrid aerogels with long-range ordered architecture and moderate density for highly thermally conductive composites. *Carbon* **126**, 119-127 (2018).
13  Wang, M. *et al.* Highly Compressive Boron Nitride Nanotube Aerogels Reinforced with Reduced Graphene Oxide. *ACS nano* **13**, 7402-7409 (2019).
14  Brown, E. *et al.* 3D printing of hybrid MoS2-graphene aerogels as highly porous electrode materials for sodium ion battery anodes. *Materials & Design* **170**, 107689 (2019).
15  Zhong, Y. *et al.* Three-dimensional MoS 2/Graphene Aerogel as Binder-free Electrode for Li-ion Battery. *Nanoscale research letters* **14**, 85 (2019).
16  Worsley, M. A. *et al.* Ultralow density, monolithic WS2, MoS2, and MoS2/graphene aerogels. *ACS nano* **9**, 4698-4705 (2015).
24